# REPRESENTATION LEARNING THROUGH CROSS-MODAL CONDITIONAL TEACHER-STUDENT TRAINING FOR SPEECH EMOTION RECOGNITION

*Sundararajan Srinivasan, Zhaocheng Huang, Katrin Kirchhoff*

Amazon AWS AI

{sundarsr, davidhzc, katrinki}@amazon.com

**ABSTRACT**

Generic pre-trained speech and text representations promise to reduce the need for large labeled datasets on specific speech and language tasks. However, it is not clear how to effectively adapt these representations for speech emotion recognition. Recent public benchmarks show the efficacy of several popular self-supervised speech representations for emotion classification. In this study, we show that the primary difference between the top-performing representations is in predicting valence while the differences in predicting activation and dominance dimensions are less pronounced. However, we show that even the best-performing HuBERT representation underperforms on valence prediction compared to a multimodal model that also incorporates text representation. We address this shortcoming by injecting lexical information into the speech representation using the multimodal model as a teacher. To improve the efficacy of our approach, we propose a novel estimate of the quality of the emotion predictions, to condition teacher-student training. We report new audio-only state-of-the-art concordance correlation coefficient (CCC) values of 0.757, 0.627, 0.671 for activation, valence and dominance predictions, respectively, on the MSP-Podcast corpus, and also state-of-the-art values of 0.667, 0.582, 0.545 on the IEMOCAP corpus.

## 1. INTRODUCTION

Speech Emotion Recognition (SER) has been gaining popularity over the last two decades due to its importance for human computer interaction, call-center, e-learning, and mental health monitoring [1]–[3]. In automated SER systems, there are two common ways to represent emotions, i.e., discrete emotion categories (e.g., anger, sadness, etc.) and continuous dimensional emotions (e.g., activation and valence, which measures how activated and positive a person feels, respectively). Recently, emotion dimensions have become popular and widely adopted, since they are more capable of capturing subtle changes in emotions and representing complex emotions, compared with the emotion categories [3], [4].

In building automated systems to recognize emotions from speech, hand-crafted features remained dominant in early days starting from prosodic tones, to brute-force functionals [3], to compact expert-driven acoustic features [5]. Over the past several years, deep learning has become the mainstay in SER [6]–[8]. However, the improvements remain limited, predominantly because emotional datasets are generally small in size, posing a major challenge for deep learning techniques to be highly effective for SER. One promising approach to circumvent this problem is the use of learned speech representations. Of particular interest is self-supervised (or unsupervised) representation learning, which leverages large unlabeled datasets to learn powerful generic speech representation using a pretext task [9]. This has yielded promising results in many downstream tasks [10], including automatic speech recognition (ASR), speaker recognition, and emotion recognition [11]–[14]. However, how best to adapt self-supervised learning to SER is still being explored.

Emotion recognition has long found to benefit from more than one modality, and therefore multimodal emotion recognition is not uncommon, including audio, video, text and physiological cues [6], [15]. Different modalities are complementary to each other, for example, audio is found to be more effective for activation, whereas video and text are found to be more effective for valence [3]. However, video and physiological signals are not always available, whereas speech and text remain more accessible and less intrusive. There have been a few studies investigating fusion of speech and text for emotion recognition [16], [17]. However, these are cumbersome to deploy, since they require access to speech-to-text systems. Moreover, novel powerful speech representations like Hidden unit BERT (HuBERT) [18] already capture long-range context, and it is not clear what benefits fusing lexical information might provide, and if it does, how we can leverage that for improving the speech representations.

In this paper, we answer this question by showing that fusion of text representations with that from HuBERT improves valence prediction. We further use this multimodal fusion model as teacher to fine-tune the audio-based representation for improving audio-based emotion recognition. We then introduce a novel metric to quantify the quality of emotion prediction, and use this to apply teacher-student training conditionally based on the quality of the teacher's predictions, leading to the state-of-the-art emotion recognition performance on MSP-Podcast.

## 2. RELATED WORK

Finding effective representation from speech has been a long-standing topic in the field of speech emotion recognition [2]. A more recent trend is to apply self-supervised learning techniques for emotion recognition [10], [11], [14]. Self-supervised learning overcomes a limitation of SER, where databases are generally small in size and thus challenging for deep learning to learn effective representation in a supervised manner. In [11], the authors showed that a pre-trained Contrastive Predictive Coding (CPC) [19] representation from 100 hours of LibriSpeech audiobook corpus is much more effective than the conventional Mel filter-bank features, leading to the best previously reported performance on the MSP-

Podcast dataset. A recent speech benchmark, SUPERB [10], studied the effectiveness of different pre-trained self-supervised representations for a wide range of speech-related applications, including emotion classification task on the IEMOCAP dataset [20]. HuBERT and Wav2Vec 2.0 [21] are among the representations that achieve best performance on the emotion task. In this work, to understand where the relative differences stem from, we compare the performance of these representations in predicting the three emotion dimensions on the larger MSP-Podcast dataset.

Many previous works have explored multimodal models for emotion recognition using both text and audio [16], [17], [22]–[24]. BERT [25] word embeddings have popularly been used to represent text for emotion recognition. In [16], high-level statistical aggregators of frame-level acoustic features, and word embeddings are input to separate LSTMs before concatenation, followed by fully connected layers for predicting activation, valence and dominance. In [22], a BERT-based system was used to generate pseudo labels (i.e., positive, neutral, negative sentiment) for datasets, which are then used to train audio systems in a semi-supervised manner. However, such a system can be limited, because the labels from a text-based system are less than reliable and the annotated corpora may not be emotionally rich. In [23], pre-trained Speech-BERT and RoBERTa were used to handle tokenized speech (via VQ-wav2vec) and text respectively before their representations are concatenated to classify emotions. Cross-modal attention is another approach to learn the interaction between audio and text [23], [24], but it does not seem to guarantee improvements over a simple concatenation [23].

In contrast, in this work, we explore what we can gain by adding text modality using BERT embeddings to already strongly contextual speech features like HuBERT, and also how this multimodal model can be leveraged to yield improved audio-only representation for emotion recognition.

## 3. PROPOSED EMOTION RECOGNITION FRAMEWORK

### 3.1. Emotion Recognition System

In this study, emotion recognition is formulated as prediction of activation, valence and dominance, i.e., a regression problem. Given a set of extracted features $\{\boldsymbol{u}^i, \boldsymbol{y}^i\}_{i=1}^{I}$, where $i \in [1, ..., I]$ represents the $i$-th speech utterance, $\boldsymbol{u}^i$ is the speech representation for utterance $i$, for $\boldsymbol{y}^i = [y_A^i, y_V^i, y_D^i]$ represents the ground truth *a*ctivation, *v*alence, and *d*ominance for utterance $i$ respectively.

Figure 1 demonstrates the network architecture of our emotion recognition systems. The speech representation module takes in audio waveform and extracts features at the frame level, $u_{A,t}$. This is then fed to a Gated Recurrent Unit (GRU) network to learn sequential information, whose last frame output is used as the utterance-level audio representation, $\boldsymbol{u}_A$. An internal speech-to-text system was applied to convert each speech utterance to a string of text, which is then input into a pre-trained BERT model [25] to obtain text representation $u_{T,w}$ for the $w$-th token. Another GRU is used to summarize information into an utterance-level text representation by using output of the last frame, $\boldsymbol{u}_T$. Joint embedding $\boldsymbol{e}_{AT}$ is obtained by concatenation of the audio and text representations, followed by a linear projection $\boldsymbol{P}_{AT}$:

$$\boldsymbol{e}_{AT} = \boldsymbol{P}_{AT}(\text{concat}[\boldsymbol{u}_A, \boldsymbol{u}_T]) \quad (1)$$

Similarly, the audio embedding $\boldsymbol{e}_A$ is obtained from a linear projection $\boldsymbol{P}_A$ of utterance-level audio representation:

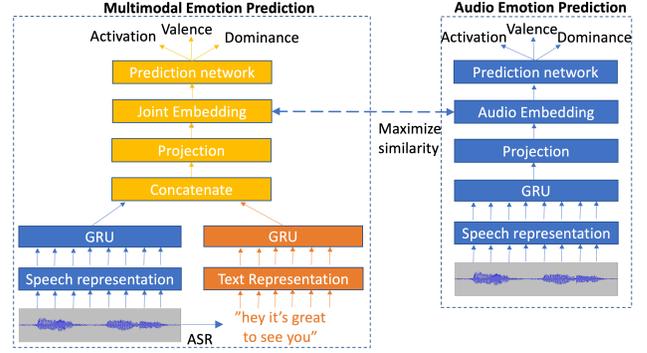

Fig. 1: *Proposed system for speech emotion prediction from audio representations, multimodal audio and text representations, and Teacher-Student training of audio model from multimodal model.*

$$\boldsymbol{e}_A = \boldsymbol{P}_A(\boldsymbol{u}_A) \quad (2)$$

The audio embedding $\boldsymbol{e}_A$ (for the audio-only model) or the joint embedding $\boldsymbol{e}_{AT}$ (for the multimodal model) is then passed through a linear prediction network that jointly predicts all the three emotion dimensions (i.e., multi-task learning).

The loss function to optimize is based on Concordance Correlation Coefficient (CCC), which is also the standard metric used to evaluate prediction performance for each of the three emotion dimensions, given predictions $\tilde{y}$ and ground truth $y$ [26]:

$$CCC = \frac{2Cov(\tilde{y}, y)}{\sigma_{\tilde{y}}^2 + \sigma_y^2 + (\mu_{\tilde{y}} - \mu_y)^2} \quad (3)$$

We also note that CCC is the product of Pearson correlation coefficient ($\rho$) and a correction term ($C_b$) that penalizes mean shift and variance scale between prediction and labels [27]:

$$CCC = \rho * C_b \quad (4)$$

We will use this fact in the next subsection. Similar to [11], [16], we used a loss function based on CCC and multi-task learning for training our emotion recognition systems.

$$\mathcal{L}_{EMO} = \sum_j \alpha_j (1 - CCC_j), j \in [A, V, D] \quad (5)$$

where $\alpha_j$ represents the weights for the three emotion dimensions and was set to equal in this study.

### 3.2. Conditional Teacher-Student Learning

To improve performance of the audio-only system, we propose to inject lexical information from the joint model. This can be achieved using teacher-student (T/S) learning, a form of transfer learning where a student model learns to mimic the output distribution of a teacher model [28]. We apply T/S learning to teach an audio-only speech emotion model to output audio embeddings as close as possible to the joint embedding from the multi-modal teacher model, by minimizing the *L2* loss between the two embeddings:

$$\mathcal{L}_{STU} = \parallel \boldsymbol{e}_{AT} - \boldsymbol{e}_A \parallel \quad (6)$$

The total loss becomes

$$\mathcal{L}_{TOTAL} = \mathcal{L}_{EMO} + \lambda * \mathcal{L}_{STU} \quad (7)$$

**Table 1.** *Comparison of CCC and Pearson's correlation coefficient ($\rho$) scores on MSP-Podcast v1.6 training set using the fusion model, showing CCC is dominated by $\rho$ for all three dimensions.*

| $CCC_A$ | $\rho_A$ | $CCC_V$ | $\rho_V$ | $CCC_D$ | $\rho_D$ |
|---|---|---|---|---|---|
| 0.763 | 0.764 | 0.687 | 0.699 | 0.715 | 0.717 |

where weight $\lambda$ is tuned on a validation set.

However, teacher prediction quality for an utterance may be poor due to several reasons, such as poor transcription, contradictory information between audio and text, or just poor model fit in that region. Motivated by [28], we condition T/S to discard utterances on which the prediction quality of the teacher is poor. However, unlike in [28], our targets are continuous variables; hence we cannot use correct classification as the criterion for T/S. Moreover, the targets and predictions can have potentially different means and variances (which is the reason for the $C_b$ correction term in equation for CCC in Eqn. 4). Hence, a naïve absolute difference between prediction and target is not a direct measure of prediction quality.

To gain more insights to derive an appropriate condition, CCC and $\rho$ values are shown in Table 1 for our multimodal teacher model on the MSP-Podcast training subset. In Table 1, it is evident that the CCC values are dominated by $\rho$ (also see Tables 2, 3 in [26] for similar observations). It is well-known in statistics that $\rho^2$ is the ratio of (least-squares) linear regression sum of squares to total sum of squares. Higher residuals (or errors) from the best linear fit of predictions to labels will lower CCC, while lower residuals will contribute to increasing CCC. Hence, we adopt these residuals as a measure of emotion prediction error.

The residual $r$ for $i$-th speech utterance can be obtained as:

$$r_j^i = w_j^T \tilde{y}_{AT,j}^i + b_j - y_j^i, j \in [A, V, D] \quad (8)$$

Where, for emotion dimension $j$, $w_j$ and $b_j$ are weight and offset parameters of the least-squares linear regression model, $\tilde{y}_{AT,j}^i$ is the prediction of the multimodal teacher model, and $y_j^i$ is the label. We discard utterances that have residuals $r_j^i$ with large magnitudes for any of the three dimensions.

## 4. EXPERIMENTS AND RESULTS

### 4.1. Datasets

We used two emotional speech corpora: the MSP-Podcast corpus (version 1.6) [29] and the IEMOCAP corpus [20], because they are among the widely adopted publicly available emotional corpora in English. MSP-Podcast has 34,280 utterances for training, 5,958 utterances for validation, and 10,124 utterances for testing, totaling 84 hours of naturalistic speech annotated in terms of emotion dimensions and categories. This corpus does not include ground-truth transcriptions. IEMOCAP has 12 hours of speech collected from 10 actors, divided into 5 sessions of dyadic interactions, with 10,039 utterances in total, including scripted and spontaneous speech. We adopted 5-fold speaker-independent cross-validation, where in each fold, one session was used for evaluation (~2,000 utterances), another for validation (~2,000 utterances), while utterances from the remaining three sessions were used for training (~6,000 utterances). This corpus has manual transcripts. In both these corpora, each utterance was rated by multiple annotators for the three emotion dimensions, and we used mean rating across all annotators as the ground-truth label.

### 4.2. Experimental Settings

We trained our model with a batch size of 32 for 50 epochs with ADAM optimizer with initial learning rate of 5e-4 and reduced the rate by a factor 0.75 when loss on validation set stops decreasing for more than two epochs. We used a 2-layer GRU with 128 hidden units for summarizing utterance-level information for each modality. We also experimented with a transformer instead of GRU, aggregating results as in [11], and we found performance was comparable, so we report results only using GRU layers. A linear projection network with output size of 128 was used to generate final embeddings, which are fed to a linear prediction network that outputs predictions for the three emotion dimensions.

For experiments that fine-tuned parameters of the speech representation model (including the teacher-student trained model), learning rate for learning representation model parameters was lower by a factor of 4 to reduce the possibility of over-training. Furthermore, we only fine-tuned the top 6 layers of the speech representation model on the training set of MSP-Podcast or IEMOCAP (5-fold). Models that yielded the best results on validation set were selected for evaluation. $\lambda$, which trades off teacher-student learning ($\mathcal{L}_{STU}$) and emotion recognition ($\mathcal{L}_{EMO}$) in Eqn. (7), was set to 30, but was observed to be relatively insensitive in the range 3-50.

For conditional T/S learning, a residual threshold of 2 standard-deviations was used, as a result of which ~12.7% of the training utterances were discarded from teacher-student $L2$ loss, though these utterances were still used to minimize CCC loss during training. We used an internal ASR system to generate transcripts from MSP-Podcast audio files, while the original ground-truth transcriptions were used for IEMOCAP. We used publicly available pre-trained models for both speech and text representations[1].

### 4.3. Results and Discussion

Table 2 presents emotion recognition performance on MSP-Podcast using pre-trained speech representations as well as the proposed (conditional) TS framework. For pre-trained models, i.e., wav2vec 2.0 and HuBERT, we adopted both the base and large models, which have been pre-trained on 960 and 60k hours of speech respectively. Firstly, we observed the same trend as in [10], as we move towards more contextualized speech features, going from Modified CPC to wav2vec 2.0 to HuBERT, CCC for valence improves considerably while activation and dominance remain relatively unchanged. To our knowledge, the valence CCC of HuBERT-large model at 0.547 already outperforms previous published numbers using audio on this dataset. Moreover, comparisons between base and large pretrained models (i.e., system 4 vs. 5 and system 6 vs. 7) suggest that a detailed representation of the acoustic space yield further benefits for valence prediction, because the large models have more parameters and were trained on much larger datasets than the base ones.

---

[1] https://huggingface.co/facebook/{wav2vec2-base-960h, wav2vec2-large-lv60, hubert-base-ls960, hubert-large-ll60k}, https://huggingface.co/bert-base-uncased

**Table 2.** *CCC scores for emotion recognition on MSP-Podcast using self-supervised learning with and without teacher-student (T/S) transfer learning. "FT": fine-tuning, "cT/S": conditional T/S learning. #n refers to the system in row n. #8 is text-only and #9 is multimodal and both are de-highlighted, rest are audio-only systems.*

|   |   | Speech Representation | $CCC_A$ | $CCC_V$ | $CCC_D$ |
|---|---|---|---|---|---|
| 1 | Baselines | Attn [33] | 0.695 | 0.307 | 0.613 |
| 2 |  | CPC [11] | 0.706 | 0.377 | 0.639 |
| 3 |  | Modified CPC | 0.736 | 0.396 | 0.662 |
| 4 |  | wav2vec 2.0 Base | 0.728 | 0.363 | 0.636 |
| 5 | Pre-trained | wav2vec 2.0 Large | 0.735 | 0.472 | 0.654 |
| 6 | Models | HuBERT Base | 0.733 | 0.485 | 0.640 |
| 7 |  | HuBERT Large | 0.752 | 0.547 | 0.674 |
| 8 |  | BERT Base | 0.317 | 0.563 | 0.290 |
| 9 | Teacher | Multimodal (#7 + #8) | 0.765 | 0.690 | 0.683 |
| 10 | FT | #6 | 0.730 | 0.533 | 0.644 |
| 11 | Our T/S | T/S (#9/#6) | 0.738 | 0.590 | 0.650 |
| 12 | framework | cT/S (#9/#6) | **0.757** | **0.627** | **0.671** |

**Table 3.** *CCC scores for emotion recognition on IEMOCAP using self-supervised learning with and without teacher-student (T/S) transfer learning. "FT": fine-tuning, "cT/S": conditional T/S learning. #n refers to the system in row n. #3 is text-only, #1 and #4 are multimodal, and are de-highlighted; rest are audio-only systems.*

|   |   | Speech Representation | $CCC_A$ | $CCC_V$ | $CCC_D$ |
|---|---|---|---|---|---|
| 1 | Baseline | Multimodal [16] (Audio+Text) | 0.594 | 0.446 | 0.485 |
| 2 | Pre-trained | HuBERT Base | 0.663 | 0.527 | 0.530 |
| 3 | Models | BERT Base | 0.463 | 0.576 | 0.442 |
| 4 | Teacher | Multimodal (#2 + #3) | 0.668 | 0.648 | 0.537 |
| 5 | Our T/S | T/S (#4/#2) | **0.667** | **0.582** | **0.545** |
| 6 | framework | cT/S (#4/#2) | 0.645 | 0.481 | 0.528 |

Secondly, the text representation (i.e., system 8) alone can outperform the best audio representations on valence prediction, though it performs poorly on activation and dominance. Fusing the two modalities (i.e., system 9) significantly improves the CCC for valence prediction, leading to the best performing system explored here. This result indicates that there is complementary information between the audio and text modalities, even when using a strong speech representation like the HuBERT model.

Thirdly, by fine-tuning HuBERT Base for emotion recognition, valence CCC improves from 0.485 to 0.533. Table 2 also shows that teacher-student training using a multimodal teacher model improved the audio-only HuBERT Base model, leading to valence CCC of 0.59 (i.e., system 11). Further, applying the proposed conditional teacher-student technique using linear prediction residual to assess the quality of prediction succeeds in improving valence CCC further to 0.627 (i.e., system 12). To our knowledge, this is the best audio-only performance on this dataset.

To assess the generalization ability of our techniques, we also evaluated performance on the IEMOCAP corpus, presented in Table 3. It is shown that models based on HuBERT representation performed favorably compared to the best performance obtained in the literature[2]. Fusing with the text modality improved valence performance considerably. Using our teacher-student approach (i.e., system 5), this gap in valence between the audio-only representation (i.e., system 2) and multimodal representation (i.e., system 4) is narrowed considerably. However, the proposed conditional teacher-student approach performed relatively poorly. This could be due to the small size of the IEMOCAP corpus (5x smaller than MSP-Podcast v1.6); removing utterances using conditional teacher-student training led to overfitting, and hence, worse performance on the held-out test set. Furthermore, the ground-truth transcripts from IEMOCAP are cleaner than those of MSP-Podcast, which protects against the possibility of poor predictions due to bad utterance transcriptions.

---

[2] The best reported baseline on IEMOCAP is [11], but their partition between train/validation/test is not speaker-independent (private correspondence with author, Bo Yang), and hence not comparable.

## 5. CONCLUSION

Speech Emotion Recognition remains challenging due to the small size of publicly available emotional corpora to date. In this paper, we tackled the problem of emotion recognition by utilizing self-supervised speech representations. Adding the text modality to the audio modality shows only minor improvement for predicting activation and dominance, but substantially improves valence prediction. This shows that relevant lexical and semantic information is not wholly captured even with strong speech representations like HuBERT. Motivated by this finding, we applied teacher-student learning to fine-tune HuBERT representation using the multimodal audio+text teacher model. We further proposed a technique to condition the teacher-student learning by deriving a quantity that inversely correlates with CCC, thus avoiding learning from poor teacher predictions. The proposed conditional T/S framework achieved state-of-the-art performance on the MSP-Podcast corpus. However, on IEMOCAP, while our T/S approach substantially improved performance of the audio-only system, conditioning T/S leads to performance degradation, possibly due to the small size of the dataset. Of particular interest is the finding that valence prediction, a longstanding challenge for speech emotion recognition, benefited considerably from more effective pre-trained self-supervised models and from injection of lexical information from the text modality. Our proposed audio-only system improved valence prediction CCC from 0.377 (previous audio-only state-of-the-art) to 0.627 on MSP-Podcast, and from 0.446 (previous multimodal state-of-the-art) to 0.582 on IEMOCAP.

As future work, we will explore using a different layer or a weighted combination of layers from the speech representation networks for emotion recognition, since different layers are known to contain different information [30]. We also believe that contrastive learning of the information shared by different modalities might contribute further to emotion recognition performance [31]. Also, of great interest is injecting lexical information into generic speech representations before fine-tuning on emotion datasets [32].